\documentclass[10pt]{amsart}
\usepackage{amsbsy,amssymb,amscd,amsfonts,latexsym,amstext,delarray,
 amsmath,graphicx}
\input{amssymb.sty}

\def\C{{\mathbb C}}
\def\Cb{{\mathbb C}}

\newcommand{\bG}{\mathbb{G}}
\renewcommand{\H}{\mathbb{H}}

\def\N{{\mathbb N}}

\renewcommand{\P}{{\mathbb P}}
\def\bP{{\mathbb P}}

\def\Q{{\mathbb Q}}
\def\R{{\mathbb R}}

\def\T{{\mathbb T}}

\def\Z{{\mathbb Z}}
\newcommand{\scr}{\mathcal}

\def\sC{{\mathcal C}}
\def\Dc{{\mathcal D}}
\def\cD{{\mathcal D}}
\def\sD{{\mathcal D}}
\def\Ec{{\mathcal E}}

\newcommand{\cF}{\scr{F}}
\def\cG{{\mathcal G}}

\def\cH{{\mathcal H}}
\def\Hc{{\mathcal H}}

\def\Mc{{\mathcal M}}
\def\M{{\mathcal M}}
\def\cM{{\mathcal M}}
\def\Nc{{\mathcal N}}
\def\Oc{{\mathcal O}}
\newcommand{\fO}{{\mathcal{O}}}
\renewcommand{\O}{{\mathcal O}}

\def\Qc{{\mathcal Q}}

\def\cT{{\mathcal T}}
\def\sT{{\mathcal T}}

\newcommand{\cU}{\scr{U}}

\newcommand{\fg}{{\mathfrak{g}}}

\def\a{\alpha}
\def\b{\beta}
\def\g{\gamma}
\def\G{\Gamma}

\def\D{\Delta}
\def\ve{\varepsilon}

\def\t{\theta}

\def\vp{\varphi}

\def\qq{{\,,\quad \forall}}

\def\ify{\infty}
\def\lgl{\langle}

\def\op{\oplus}
\def\ot{\otimes}

\def\part{\partial}
\def\rgl{\rangle}
\def\sbs{\subset}

\def\ra{\rightarrow}

\def\text{\hbox}
\def\Aut{\mathop{\rm Aut}\nolimits}

\def\End{\mathop{\rm End}\nolimits}
\def\Ext{{\rm Ext}}
\def\Gal{{\rm Gal}}
\def\GL{{\rm GL}}
\def\Gr{{\rm Gr}}
\def\Hom{\mathop{\rm Hom}\nolimits}

\def\Ker{\mathop{\rm Ker}\nolimits}
\def\Lie{{\rm Lie}}

\def\Sp{{\rm Spec}}
\def\Spec{{\rm Spec}}

\def\Tr{{\rm Tr}}

\newcommand{\ie}{{\it i.e.\/}\ }
\newcommand{\eg}{{\it e.g.\/}\ }
\newcommand{\cf}{{\it cf.\/}\ }

\newtheorem{thm}{Theorem}[section]
\newtheorem{prop}[thm]{Proposition}

\numberwithin{equation}{section}
\title{Quantum Fields and Motives}
\author{Alain Connes}
\author{Matilde Marcolli}
\address{A.~Connes: Coll\`ege de France \\
3, rue d'Ulm \\ Paris, F-75005 France}
\email{alain\@@connes.org}
\address{M.~Marcolli: Max--Planck Institut f\"ur Mathematik  \\
Vivatsgasse 7 \\
Bonn, D-53111 Germany}
\email{marcolli\@@mpim-bonn.mpg.de}
\date{}

\begin{document}
\maketitle


\section{Renormalization: particle physics and Hopf algebras}

The main idea of renormalization is to correct the original
Lagrangian of a quantum field theory by an infinite series of
counterterms, labelled by the Feynman graphs that encode the
combinatorics of the perturbative expansion of the theory. These
counterterms have the effect of cancelling the ultraviolet
divergences. Thus, in the procedure of perturbative
renormalization, one introduces a counterterm $C(\G)$ in the
initial Lagrangian for every divergent one particle irreducible
(1PI) Feynman diagram $\Gamma$. In the case of a {\em
renormalizable} theory, all the necessary counterterms $C(\G)$ can
be obtained by modifying the numerical parameters that appear in
the original Lagrangian. It is possible to modify these parameters
and replace them by (divergent) series, since they are not
observable, unlike actual physical quantities that have to be
finite. One of the fundamental difficulties with any
renormalization procedure is a systematic treatment of nested and
overlapping divergences in multiloop diagrams.

\medskip
\subsection*{Dimensional regularization and minimal
subtraction}\hfill\medskip

One of the most effective renormalization techniques in quantum field
theory is dimensional regularization (DimReg). It is widely used in
perturbative calculations. It is based on an analytic continuation of
Feynman diagrams to complex dimension $d\in \C$, in a neighborhood of
the integral dimension $D$ at which UV divergences occur. For the
complex dimension $d\to D$, the analytically continued integrals
become singular and the expression admits a Laurent series
expansion. Thus, within the framework of dimensional regularization,
one can implement a renormalization by minimal subtraction, where the
singular part of the Laurent series in $z=d-D$ is subtracted at each
order in the loop expansion. This renormalization method (DimReg plus
minimal subtraction) was developed by `t Hooft and Veltman \cite{tHV},
who applied it to one-loop calculations in scalar electrodynamics,
discussed the problem of overlapping divergences, the Ward identities,
the case of theories with fermions, and anomalies.  The method has
since been applied widely to perturbative calculations and it quickly
became the standard regularization and renormalization method for
nonabelian gauge theories and the standard model.

\medskip
\subsection*{Hopf algebras and the combinatorics of
renormalization}\hfill\medskip

The modern viewpoint on combinatorics, which unfolded in the 70s
around the systematic and rigorous restructuring of its foundations
advocated by Giancarlo Rota, showed how algebraic structures such as
coalgebras, bialgebras, and Hopf algebras govern elaborate
combinatorial phenomena (\cf \cite{JoRo},
\cite{Ro}, \cite{Sch}). The reason why such algebraic structures are
naturally present lies in the fact that combinatorial objects
tend to admit decomposition laws that reduce them to simpler pieces.
Such laws are the source of coproduct rules. This principle was
illustrated by many examples of incidence Hopf algebras arising from
classes of graphs and matroids. The typical situation is families of
finite graphs, closed under disjoint union and taking vertex induced
subgraphs. These admit a coproduct of the form
\begin{equation}\label{coprodcomb}
 \Delta(\Gamma)= \sum_{W\subseteq V} \gamma_W \otimes
\gamma_{V-W},
\end{equation}
where $V$ is the set of vertices of $\Gamma$ and $\gamma_W$ is the
induced subgraph on a set of vertices $W\subseteq V$. Results from
Hopf algebras in combinatorics were used, for instance, to study
graph coloring problems.

\smallskip

During 1960s and 1970s Quantum Field Theory underwent a season of
extraordinary theoretical development.
The detailed knowledge theoretical physicists gained on the subject
not only made it into something of an art, but refined it into a
highly sophisticated instrument, capable of producing
theoretical predictions that, to this day, match experiments with
unprecedented precision. Renormalization plays a central role in
the quantum theory of fields, in as it provides a consistent scheme
for extracting from divergent expressions finite values that can be
matched to physically observed quantities. Various renormalization
schemes can be implemented (though here we will be
concerned only with the ``dimensional regularization and minimal
subtraction'' scheme described above). A renormalization scheme produces an
extremely elaborate combinatorial recipe that accounts for structuring
of subgraphs in a hierarchy of subdivergences and counterterms.
Perturbative renormalization hence appears as one of the most elaborate
combinatorial recipes imposed on us by nature.

\smallskip

Conceptually, the crucial issue in the combinatorics of perturbative
renormalization is a scheme that accounts for subdivergences. This is
achieved by a {\em forest formula}, which assigns to a graph $\Gamma$
a formal expression where the subdivergences have been dealt with through
inductively defined counterterms.
Subtraction of the corresponding
counterterm from this formal expression finally yields the
renormalized value for $\Gamma$.
The definition of such formal
expressions, as we discuss more in detail below, is related to
decomposing a graph by extracting all possible divergent subgraphs
$\gamma\subset \Gamma$ and considering corresponding graphs
$\Gamma/\gamma$ obtained by collapsing $\gamma\subset \Gamma$ to a
single vertex. Such decomposition is more complicated than
those derived from incidence relations on graphs in many combinatorial
problems, as it is adapted to the specific divergences of the physical
theory and has to take into account other data like the distribution
of external momenta. Still, one can see a suggestive analogy between
the type of decomposition expressed by coproducts \eqref{coprodcomb} and the
decomposition
\begin{equation}\label{forest}
\Delta(\Gamma)= \sum_{\gamma \subseteq \Gamma} \gamma \otimes \Gamma/\gamma
\end{equation}
in a sum over divergent subgraphs, which underlies the
combinatorics of the forest formulae.
It was the seminal work of Dirk Kreimer \cite{DK1} in 1997 that
paved the way to a conceptual mathematical
formulation of perturbative renormalization, precisely by encoding the
complicated combinatorics of forest formulae via a coproduct
\eqref{forest} and identifying the Hopf algebra that governed the
renormalization procedure.

\smallskip

The extraction of a renormalized value from divergent Feynman
integrals was related in \cite{DK1} and \cite{CK3} to the antipode
in the Hopf algebra. However, the precise formula for the
renormalized value given by the BPHZ procedure (\cite{BP},
\cite{Hepp}, \cite{Zi}) requires a further operation that twists
the antipode, which, in this formulation, is not given directly in
terms of the Hopf algebra structure. The main conceptual
breakthrough in the understanding of the renormalization procedure,
that fully reconciles it with the Hopf algebra structure,
was then obtained in a later stage of development of the
Connes--Kreimer theory of perturbative renormalization,
\cite{CK1}, \cite{CK2}, where the BPHZ recursive formulae (see
\eqref{BPprep}, \eqref{counter}, \eqref{Rgamma} below) are
described in terms of the Birkhoff factorization of loops. We
shall return to this point in Section \ref{SectBirk}.

\smallskip

Given the  state of affairs in combinatorics and in quantum field
theory around the late '70s, it may seem surprising that the
pursuit of a conceptual mathematical interpretation of the
procedure of perturbative renormalization had to wait, as it did,
until the late 1990s. One should keep in mind though that, during
the 1970s, mathematicians and physicists were maximally apart. The
tendency among physicists was to shift the emphasis heavily
towards deriving efficient computational recipes at the expense of
conceptual understanding, the latter being often dismissed as a
mere exercise of pedantry. This position, though justifiable in
developing a theoretical apparatus that could be continuously
tested against experiments, had the effect of alienating
mathematicians. While quantum mechanics stimulated and in turn
benefited from a lot of advancements in modern mathematics
(operator algebras, functional analysis), mathematicians shied
away from quantum field theory, which they perceived as ill
founded, riddled with inevitable divergences, and governed by
obscure hands-on recipes. In more recent times, mathematicians and
theoretical physicists found a renewed harmony of language, but
this happened mostly in the context of string theory. This,
however, bypasses many of the crucial problems posed by quantum
field theory, by proposing a large restructuring of the
foundations of high energy physics, which at present still awaits
experimental confirmation. Thus, in particular, the new
developments left pretty much untouched the problem of a
conceptual understanding of the foundations of quantum field
theory. Of course, there were at various times attempts to
axiomatize quantum field theory in a way that would be palatable
for mathematicians (algebraic and constructive quantum field
theory, for instance). Such attempts unfortunately fell short of
incorporating the full complexity of quantum field theory,
especially with respect to the issue of perturbative
renormalization. On the other hand, at present perturbative
quantum field theory still remains the most accurate instrument
for theoretical predictions in elementary particle physics and
this impressive agreement between theory and nature calls for the
best possible conceptual understanding of its foundational
principles.

\medskip
\subsection*{Bogoliubov--Parasiuk preparation}\hfill\medskip

The Bogoliubov--Parasiuk  preparation, or BPHZ method (for
Bogoliubov--Parasiuk--Hepp--Zimmermann, \cite{BP}, \cite{Hepp},
\cite{Zi}) accounts for the presence of subdivergences,
simultaneously taking care of the problem of the appearance of
non-local terms and the organization of subdivergences via an inductive
procedure.

\smallskip

The BP preparation of a graph $\Gamma$, whose divergent integral we
denote by $U(\Gamma)$, is given by the formal expression
\begin{equation}\label{BPprep}
\overline{R}( \G) = U(\G) + \sum_{\g \sbs \G} C(\g) U( \G / \g),
\end{equation}
where the sum is over divergent subgraphs. The $C(\g)$ are inductively
defined counterterms, obtained (in the minimal subtraction scheme)
by taking the pole part (here denoted by $T$) of the Laurent expansion
in $z=d-D$ of a divergent expression,
\begin{equation}\label{counter}
C(\G) =  -T(\overline{R}( \G)) = -T\left(U(\G) + \sum_{\g \sbs \G}
 C(\g) U( \G /
 \g)\right).
\end{equation}
The renormalized value of $\Gamma$ is then given by the formula
\begin{equation}\label{Rgamma}
R(\G) =  \overline{R}( \G) +C(\G) =U(\G)  +C(\G) + \sum_{\g \sbs \G}
 C(\g) U( \G /
 \g).
\end{equation}

\smallskip

Before continuing with the physics, we need to introduce some
algebraic notions that will be useful in the rest of the paper.

\medskip
\subsection*{Hopf algebras and affine group schemes}\hfill\medskip

While affine schemes are the geometric manifestation of commutative
algebras, affine group schemes are the geometric counterpart of commutative
Hopf algebras. The theory of affine group schemes is developed in SGA
3 \cite{SGA3}.

\smallskip

Consider a commutative Hopf algebra $\Hc$ over a field $k$, which
we assume here of characteristic zero. Thus, $\Hc$ is a
commutative algebra with unit over $k$, endowed with a (not
necessarily co-commutative) coproduct $\Delta: \Hc \to
\Hc\otimes_k \Hc$, a counit $\ve: \Hc \to k$, which are
$k$-algebra morphisms and an antipode $S: \Hc \to \Hc$ which is a
$k$-algebra antihomomorphism. These satisfy the ``co-rules''
\begin{equation}\label{corules}
 \begin{array}{ll}
(\Delta \otimes id)\Delta = (id\otimes \Delta)\Delta & : \Hc \to
\Hc\otimes_k \Hc \otimes_k \Hc , \\[2mm]
 (id\otimes \ve)\Delta =id = (\ve \otimes id)\Delta & : \Hc \to \Hc , \\[2mm]
 m (id \otimes S)\Delta = m (S\otimes id) \Delta = 1\,\ve & : \Hc \to \Hc,
\end{array} \end{equation}
where we used $m$ to denote multiplication in $\Hc$.

\smallskip

One then lets $G=\,{\rm Spec}\,\Hc$ be
the set of prime ideals of the commutative $k$-algebra $\Hc$, with the
Zariski topology. The Zariski topology is too coarse
to fully recover the ``algebra of coordinates'' $\Hc$ from the
topological space $\Sp(\Hc)$, but one recovers it through the
data of the structure sheaf, \ie by considering global sections of
the ``sheaf of functions'' on $\Sp(\Hc)$.

\smallskip

Since $\Hc$ is a commutative $k$-algebra, $G=\Sp(\Hc)$ is an
affine scheme over $k$, while the additional structure given by
the co-rules \eqref{corules} endow $G=\Sp(\Hc)$ with a product
operation, a unit, and an inverse.

\smallskip

More precisely, one can view such $G$ as a functor that associates
to any unital commutative algebra $A$ over $k$ a group $G(A)$,
whose elements are the $k$-algebra homomorphisms
$$
\phi \,: \Hc \to A\,,\quad \phi(x\,y)= \phi(x) \phi(y) \qq
x,y\in \Hc\,, \quad\phi(1)=1\,.
$$
The product in $G(A)$ is given as the dual of the coproduct, by
 \begin{equation} \label{dualprod}
 \phi_1\,\star\,\phi_2\,\,(x)=\,\langle  \phi_1\otimes
\phi_2\,,\;\Delta(x)\rangle\,.
 \end{equation}
The inverse and the unit of $G(A)$ are determined by the antipode
and the co-unit of $\Hc$. The co-rules imply that these operations
define a group structure on $G(A)$. The resulting covariant
functor
$$
A \,\rightarrow G(A)
$$
from commutative algebras to groups is representable (in fact by
$\Hc$). The functor $G$ obtained in this way is called an {\em
affine group scheme}. Conversely, any covariant representable
functor from the category of commutative algebras over $k$ to
groups, is an affine group scheme $G$, represented by a
commutative Hopf algebra, uniquely determined up to canonical
isomorphism.

\smallskip

Some simple examples of affine group schemes:
\begin{itemize}
\item
The multiplicative group $G=\bG_m$ is the affine group scheme obtained
from the Hopf algebra $\Hc=k[t,t^{-1}]$ with coproduct $\Delta(t)=t\otimes
t$.
\item
The additive group $G=\bG_a$ corresponds to the Hopf algebra
$\Hc=k[t]$ with coproduct $\Delta(t)=t\otimes 1 + 1 \otimes t$.
\item
The affine group scheme $G=\GL_n$ corresponds to the Hopf
algebra $$\Hc=k[x_{i,j},t]_{i,j=1,\ldots,n} / \det(x_{i,j})t-1,$$
with coproduct $\Delta(x_{i,j})= \sum_k x_{i,k}\otimes x_{k,j}$.
\end{itemize}

\smallskip

The latter example is quite general. In fact, if $\Hc$
is finitely generated as an algebra over $k$, then the corresponding
affine group scheme $G$ is a linear algebraic group over $k$, and can
be embedded as a Zariski closed subset in some $\GL_n$.
Moreover, in the more general case, one can find a collection $\Hc_i\subset \Hc$
of finitely generated algebras over $k$ such that
$\Delta(\Hc_i)\subset \Hc_i\otimes \Hc_i$, $S(\Hc_i)\subset \Hc_i$,
for all $i$, and such that, for all $i,j$ there exists a $k$ with
$\Hc_i \cup \Hc_j \subset \Hc_k$, and $\Hc=\cup_i \Hc_i$.
In this case, one obtains linear algebraic groups $G_i=\Sp(\Hc_i)$ such
that
\begin{equation}\label{Gprojlim}
 G=\varprojlim_i G_i.
\end{equation}
Thus, in general, an affine group scheme is a projective limit of
linear algebraic groups. If the $G_i$ are unipotent, then $G$ is a
pro-unipotent affine group scheme.

\smallskip

The Lie algebra $\fg(k)=\Lie\, G(k)$ is given by
the set of linear maps $ L\,:\Hc \to k$ satisfying
\begin{equation}\label{Liescheme}
L(X\,Y)=\, L(X)\,\ve(Y) +\, \ve(X)\, L(Y)\,,\quad \forall X\,,Y
\in \Hc\,,
\end{equation}
where $\ve$ is the co-unit of $\Hc$, playing the role of the unit
in the dual algebra. Equivalently, $\fg=\hbox{Lie} \ G$ is a
covariant functor
\begin{equation}\label{LieGA}
A \,\rightarrow \fg(A)\,,
\end{equation}
from commutative $k$-algebras to Lie algebras, where $\fg(A)$ is the
Lie algebra of linear maps $ L\,:\Hc \to A$ satisfying
\eqref{Liescheme}.

\medskip
\subsection*{Hopf algebra of Feynman graphs and
diffeographisms}\hfill\medskip

The Kreimer Hopf algebra of \cite{DK1} is based on
rooted trees, which organize the hierarchy of subdivergences in
a given graph. The Hopf algebra depends on the particular physical theory
$\cT$ through the use of trees whose vertices are decorated by the
divergence free Feynman graphs of the theory (\cf \cite{DK1} \cite{CK3}).
In the work of Connes--Kreimer \cite{CK1} this Hopf algebra was
refined to a Hopf algebra $\Hc(\sT)$, also dependent on the physical
theory $\sT$ by construction, which is directly defined in terms of
Feynman graphs.

\smallskip

The CK Hopf algebra is the free commutative algebra over $k=\C$
generated by one particle irreducible (1PI) graphs
$\Gamma(p_1,\ldots,p_n)$, where $\Gamma$ is not a tree. A graph
$\Gamma$ is 1PI if it cannot be disconnected by the removal of a
single edge. Here one considers graphs endowed with external
momenta $(p_1,\ldots,p_n)$. To account for this external structure
one considers distributions $\sigma\in C_c^{-\infty}(E_\Gamma)$
for $$ E_\Gamma=\,\left\{ (p_i)_{i=1 , \ldots , N} \ ; \
 \sum \, p_i = 0 \right\}, $$
and the symmetric algebra $\Hc=\,S(C_c^{-\infty}(\cup E_\Gamma))$,
with $\cup E_\Gamma$ the disjoint union.

\smallskip

\smallskip

The coproduct is given by a formula that reflects the BP
preparation \eqref{BPprep}, namely, it is given on generators by
the expression
\begin{equation}\label{CKcoprod}
  \Delta(\G) = \G \ot 1 + 1 \ot \G + \sum_{\g \sbs \G} \g_{(i)} \ot \G /
 \g_{(i)}.
\end{equation}
Here the sum is over divergent subgraphs $\gamma \sbs \G$ and $\G
/\g$ denotes the graph obtained by contracting $\g$ to a single
vertex. In \eqref{CKcoprod} the notation $\gamma_{(i)}$ accounts
for the fact that one has to specify how to assign the external
structure to $\gamma$, depending on the type of the corresponding
vertex in $\G /
 \g_{(i)}$, \cf \cite{CK1}.

\smallskip

Up to passing to the Hopf subalgebra constructed on 1PI graphs
with fixed external structure, one can reduce to a Hopf algebra
$\Hc(\sT)$ that is finite dimensional in each degree, where the
degree is defined on 1PI graphs by the loop number. There is an
affine group scheme associated to this Hopf algebra $\Hc(\sT)$.
This is called the group of {\em diffeographisms} $G={\rm
Difg}(\sT)$ of the physical theory. It is a pro-unipotent affine
group scheme.

\smallskip

The reason for the terminology lies in the fact that ${\rm
Difg}(\sT)$ has a close relation to the group of formal
diffeomorphisms of the complexified coupling constants of the
theory. In the simplest case this group is the group ${\rm
Diff}(\C)$ of formal diffeomorphisms of the complex line tangent
to the identity. The latter corresponds to the Hopf algebra
$\Hc_{\text{diff}}$ whose generators $a_n$ are obtained by writing
formal diffeomorphisms as $ \varphi(x) = x + \sum_{n\geq 2}
a_n(\varphi)\, x^n $, and with coproduct $\lgl \D a_n  \, , \,
\vp_1 \ot \vp_2   \rgl = a_n (\vp_2 \circ \vp_1)$. A Hopf algebra
homomorphism is obtained by writing the effective coupling
constant as a formal power series $g_{{\rm eff}}(g) = g +
\sum_{n\geq 2} \alpha_n \, g^n$,
 where all the coefficients $\alpha_n$ are finite linear combinations
of products of graphs, $\alpha_{n} \in \Hc$, for all $n\geq 1$ and
mapping  $a_n \mapsto \alpha_n$, \cf \cite{CK1}.

\medskip
\section{Birkhoff factorization and
renormalization}\label{SectBirk}

Suppose given a complex Lie group  $G(\C)$ and a smooth simple
curve $C\sbs \P^1 (\C)$, with $C^\pm$ the two complementary
regions, with $\infty\in C^-$. For a given loop $\g : C \to
G(\C)$, the problem of Birkhoff factorization asks whether there
exist holomorphic maps $\g_{\pm} : C_{\pm} \to G(\C)$, such that
\begin{equation}\label{Birk}
\g \, (z) = \g_- (z)^{-1} \, \g_+ (z) \qquad z \in C .
\end{equation}

\smallskip

This procedure of factorization of Lie group valued loops
became well known in algebraic geometry because of its use in the
Grothendieck--Birkhoff decomposition \cite{Gro1} of holomorphic vector
bundles on the sphere $\P^1(\C)$.
In this case, the Lie group is $\GL_n(\C)$ and
a weaker form of \eqref{Birk} holds, whereby loops factor as
\begin{equation}\label{BirkL}
\gamma(z) = \gamma_-(z)^{-1}\, \lambda(z)\, \gamma_+(z),
\end{equation}
where $ \lambda(z)$ is a diagonal matrix with entries
$(z^{k_1},z^{k_2},\cdots,z^{k_n})$.
The Grothendieck--Birkhoff decomposition hence states that a
holomorphic vector bundle on $\P^1(\C)$ can be described as
$E = L^{k_1} \op \ldots \op L^{k_n}$,
where the line bundles $L^{k_i}$ have Chern class $c_1 \, (L^{k_i})
=k_i$. This corresponds to the Birkhoff decomposition \eqref{Birk}
when $c_1 \, (L^{k_i}) = 0$.

\smallskip

{}From a more analytic viewpoint (\cf \eg \cite{Boj}), the Birkhoff
factorizations \eqref{Birk} or \eqref{BirkL} can be
viewed as a (homogeneous) {\em transmission problem}, which can be
formulated in terms of systems of singular integral equations, with
various regularity assumptions. Such transmission problems can be
recast in the context of the theory of Fredholm pairs, obtained by
considering the spaces of boundary values, on a simple closed curve
$C$, of sections of holomorphic vector bundles on $\P^1(\C)$.

\bigskip

\medskip
\subsection*{BPHZ as a Birkhoff factorization}\hfill\medskip

One of the key results of the Connes--Kreimer theory of perturbative
renormalization \cite{CK1} \cite{CK2} is a reformulation of the BPHZ
procedure as a Birkhoff factorization in the pro-unipotent Lie group
$G(\C)$ associated to the affine group scheme $G={\rm Difg}(\sT)$.

\smallskip

Unlike the case of $\GL_n$, where the Birkhoff decomposition
\eqref{Birk} only holds when $k_i=0$, in the case of interest for
renormalization one always has a factorization \eqref{Birk}. This
follows from a result of Connes--Kreimer, which we recall in
Proposition \ref{BirkHopf} below. For the general case where $G$
is the pro-unipotent affine group scheme of a Hopf algebra that is
graded in positive degree and connected, the result shows that a
factorization of the form \eqref{Birk} always exists. The result,
in fact, provides an explicit recursive formula, in Hopf algebra
terms, which determines both terms in the factorization.

\smallskip

In this setup, the Lie group $G(\C)$ is the set of complex points of
an affine group scheme $G$, whose commutative Hopf algebra $\cH$ is
graded in positive degrees $\cH=\cup_k \cH_k$ and connected (\ie the
only elements of degree $0$ in $\cH$ are the scalars).

\smallskip

We let $K=\C(\{ z \})$ be the field of Laurent series convergent
in some neighborhood of the origin (\ie germs of meromorphic
functions at the origin) and $\O=\C\{ z \}$ be the ring of
convergent power series, and we let $\Qc=\,z^{-1}\,\C([z^{-1}])$,
with $\tilde\Qc=\C([z^{-1}])$ the corresponding unital ring. Then
a loop $\gamma\,:C\to G$, for $C$ an infinitesimal circle around
the origin, is equivalently described by a homomorphism $\phi: \cH
\to K$, \ie by a point in $G(K)$. Because the group structure on
$G$ corresponds to the co-rules of the Hopf algebra $\cH$, the
product of loops $\gamma(z)=\gamma_1(z)\,\gamma_2(z)$, for $z\in
C$, corresponds to $\phi= \phi_1\star\phi_2$ (dual to the
coproduct in $\cH$) and the inverse $ z\mapsto \gamma(z)^{-1}$ to
the antipode $\phi \circ S$.

\smallskip

For $z=0 \in C^+$, the
condition that the loop $\gamma$ extends to a holomorphic function
$\gamma\,:P_1 (\Cb)\backslash\{ 0 \}\to G$ is equivalent to the
condition that the homomorphism $\phi$ lies in $G(\tilde\Qc)=\{
\phi\,,\phi({\mathcal H})\subset\;\tilde\Qc \}$, while the condition
that $\gamma(0)$ is finite translates in the condition that
$\phi$ belongs to $G(\O)=\{ \phi\,, \phi({\mathcal
 H})\subset\;\O \}$. The
normalization condition $\gamma(\infty)=1$ translates
algebraically into the condition $\ve_-\circ \phi=\,\ve$, where
$\ve_-$ is the augmentation in the ring $\tilde\Qc$ and $\ve$ is
the augmentation (co-unit) of $\Hc$. This dictionary shows how
interpreting affine group schemes as functors of unital
commutative algebras to groups provides a very convenient language
in which to reformulate the problem of Birkhoff factorization.

\smallskip

\begin{prop}\label{BirkHopf} {\em (\cite{CK1})}
Let $\cH$ be a Hopf algebra that is graded in
positive degree and connected, and $G$ the corresponding affine group
scheme. Then any loop $\gamma: C \to G(\C)$ admits a Birkhoff
factorization \eqref{Birk}. An explicit recursive formula for the
factorization is given, in terms of the corresponding homomorphism
$\phi: \cH \to \C(\{ z \})$, by the expressions
\begin{equation}\label{Hbirkhoff1}
 \phi_-(X)=-T\left(\phi(X)+\sum\phi_-(X^\prime)
 \phi(X^{\prime\prime}) \right)
 \end{equation}
 and
 \begin{equation}\label{Hbirkhoff2}
 \phi_+(X)=\phi(X)+\phi_-(X)+\sum\phi_-(X^\prime)
 \phi(X^{\prime\prime}),
 \end{equation}
where $T$ is the projection along $\O$ to the augmentation
ideal of $\tilde\Qc$ (taking the pole part), and $X'$
and $X''$ denote the terms of lower
 degree in the coproduct
 $\Delta(X)= X \otimes 1 + 1 \otimes X + \sum X^\prime
 \otimes X^{\prime\prime}$,
 for $X \in{\mathcal H}$.
\end{prop}

Applied to the Hopf algebra $\cH(\sT)$ of Feynman graphs, with $G={\rm
Difg}(\sT)$, the formulae \eqref{Hbirkhoff1} and \eqref{Hbirkhoff2}
yield the counterterms \eqref{counter} and the renormalized values \eqref{Rgamma}
in the BPHZ renormalization procedure.

\medskip
\subsection*{Mass parameter, counterterms, and the
renormalization group}\label{MassSect}\hfill\medskip

In DimReg, when analytically continuing the Feynman graphs to complex
dimension, in order to preserve the dimensionality of the integrand in
physical units, one needs to replace the momentum space integration
$d^{D-z}k$ by $\mu^z d^{D-z} k$, where $\mu$ is a mass parameter, so
that the resulting quantity has the correct dimensionality of
(mass)$^D$. This introduces a dependence on the parameter $\mu$ in the
loop $\gamma_\mu(z)$ describing the unrenormalized theory. The
behavior of a renormalizable theory under rescaling of the mass
parameter $\mu \mapsto e^t\mu$, for $t\in \R$, was analyzed in
\cite{tH}.

\smallskip

An important result, which will play a crucial role in our geometric
formulation in Section \ref{GalSect}, is that {\em the counterterms do
not depend on the mass parameter} $\mu$ (\cf
\cite{Collins} \S 5.8 and \S 7.1). This result translates in terms of the
Birkhoff factorization to the condition that the negative part
$\gamma_{\mu^-}(z)$ of the factorization $\g_{\mu} (z) = \g_{\mu^-}
(z)^{-1} \, \g_{\mu^+} (z)$ satisfies
\begin{equation}
 \frac{\partial}{\partial \mu} \, \g_{\mu^-} (z) = 0 \, . \label{gammamu-}
\end{equation}

\smallskip

The effect of scaling the mass parameter on the loop $\gamma_\mu(z)$
is instead described by the action of the 1-parameter group of
automorphisms generated by the grading by loop number. Namely, if
$\theta_t$ denotes the 1-parameter group with infinitesimal generator
$\frac{d}{dt}\theta_t|_{t=0}=Y$, where $Y$ is the grading by loop
number, we have
\begin{equation}\label{thetagamma}
\gamma_{e^t\mu}(z) =\theta_{tz}(\gamma_\mu(z)) , \ \ \ \  \forall t\in
\R,
\end{equation}
and for all $z$ in an infinitesimal punctured neighborhood $\Delta^*$
of the origin $z=d-D=0$.

\smallskip

A well known but unpublished result of `t Hooft shows that the counterterms
in a renormalizable
quantum field theory can be reconstructed from the beta function of
the theory. In the context of the Connes--Kreimer theory of
perturbative renormalization, this can be seen in the following way.

\smallskip

The beta function here is lifted from the space of the coupling
constants of the theory to the group of Diffeographisms, namely, it
can be regarded as an element in the Lie algebra $\Lie G$ satisfying
\begin{equation}\label{beta}
\b = Y \, {\rm Res} \, \g ,
\end{equation}
where $Y$ is the grading by loop number, and the residue of $\gamma$
is given by
\begin{equation}\label{resgamma}
{\rm Res}_{z = 0}^{} \g = - \left( \frac{\partial}{\partial u} \,
  \g_- \left( \frac{1}{u} \right) \right)_{u=0}.
\end{equation}
The beta function is the infinitesimal generator $\b=  \frac{d}{dt}
{\bf rg}_t
|_{t=0}$ of the renormalization group
\begin{equation}\label{rengroup}
{\bf rg}_t ={\rm lim}_{z \ra 0} \:\g_- (z) \, \t_{t z} (\g_-
 (z)^{-1}).
\end{equation}
Correspondingly, the renormalized value, that is, the finite value
$\g_{\mu}^+(0)$ of the Birkhoff
decomposition satisfies the equation
\begin{equation}\label{renaction}
  \mu \frac{\partial}{\partial \mu}
 \,\g^+_{\mu}(0)=\, \b\,\g^+_{\mu}(0)\,.
\end{equation}

\smallskip

A strong form of the `t Hooft relations, deriving the counterterms from the beta
function, is given by the following result.

\begin{prop}\label{thooftrel} {\em (\cite{CK2})}
The negative part of the Birkhoff factorization $\gamma_-(z)$ satisfies
\begin{equation}\label{gammaminussum}
\gamma_-(z)^{-1} = 1 + {\displaystyle
\sum_{n=1}^{\ify}} \ \frac{d_n}{z^n},
\end{equation}
where the coefficients $d_n$ are given by iterated integrals
\begin{equation}\label{dncond}
 d_n = \int_{s_1 \geq s_2 \geq \cdots \geq s_n \geq 0} \t_{-s_1}
(\b) \, \t_{-s_2} (\b) \ldots \t_{-s_n} (\b) \, \, ds_1 \cdots
ds_n \, .
\end{equation}
\end{prop}

The result can be formulated (\cf \cite{CK2}) as a scattering formula
\begin{equation}\label{scattering}
\g_- (z) = \lim_{t \ra \ify}
 e^{-t \left( \frac{\b}{z} + Z_0 \right)} \, e^{t Z_0},
\end{equation}
where $Z_0$ is the additional generator of the Lie algebra of $G
\rtimes_\theta \bG_a$, satisfying
\begin{equation}
 [Z_0 , X] = Y(X) \qquad \forall \, X \in \Lie\, G \, . \label{LieGstar}
\end{equation}

\smallskip

This form of the `t Hooft relations and the explicit formula
\eqref{dncond} in terms
of iterated integrals are the starting point for our formulation of
perturbative renormalization in terms of the Riemann--Hilbert
correspondence and for the relation to motivic Galois theory.

\smallskip

Before
continuing with a more detailed account of these topics, we give an
introductory tour of some ideas underlying the
theory of motives and the Riemann--Hilbert correspondence, that we
will need in order to introduce the main result of \cite{cmln}.

\section{The yoga of motives: cohomologies as avatars}\label{MotSect}

There are several possible cohomology theories that can be applied to
algebraic varieties. Over a field $k$ of characteristic zero one has
de Rham cohomology $H_{dR}^\cdot(X)=\H^\cdot(X,\Omega_X^\cdot)$, defined
in terms of sheaves of differential forms, and Betti cohomology
$H^\cdot_B(X,\Q)$, which is a version of singular homology for $\sigma
X(\C)$, for an embedding $\sigma: k\hookrightarrow \C$. These are
related by the periods isomorphism
$$ H^i_{dR}(X,k)\otimes_\sigma \C \cong H^i_B (X,\Q)\otimes_\Q \C. $$
Over a perfect field of positive characteristic there is also crystalline
cohomology, while in all characteristics one can consider \'etale
cohomology given by finite dimensional $\Q_\ell$-vector spaces
$H^i_{et}(\bar X, \Q_\ell)$, where $\bar X$ is obtained by
extension of scalars to an algebraic closure $\bar k$, and $\ell\neq
{\rm char}\, k$. In the smooth projective case, these have the
expected properties of Poincar\'e duality, K\"unneth isomorphisms,
etc. Moreover, \'etale cohomology provides interesting $\ell$-adic
representations of $\Gal(\bar k/k)$. There are comparison
isomorphisms
$$ H^i_B(X,\Q)\otimes_\Q \Q_\ell \cong H^i_{et} (\bar X,\Q_\ell). $$

\smallskip

The natural question is then what type of information, such as
maps or operations on one cohomology, can be transferred to the other
ones. This gave rise to the idea, proposed by Grothendieck, of the
existence of a ``universal cohomology theory'' with realization
functors to all the known cohomology theories for algebraic
varieties. He called this the theory of {\em motives}.

\smallskip

A metaphor \cite{Gro3} justifying the terminology is provided by
music scores, some of which (such as Bach's ``Art of the fugue")
are not written for any particular instrument. They are just the motive,
which in turn can be realized on different musical instruments.
Another powerful metaphor is provided by the notion of avatar in Hindu
philosophy, which expresses the idea of a single
entity manifesting itself in manifold incarnations (the ten
avatars of Vishnu).

\smallskip

We will present here only a very short overview of some ideas and results about
motives, following \cite{De3}, \cite{Man2}, \cite{Se1}, and \cite{Blo},
\cite{dg}, \cite{Gon}, \cite{Le}.
We start first by recalling some general algebraic formalism we will
need in the following.

\medskip
\subsection*{Tannakian categories}\hfill\medskip

The basis for a Galois theory of motives lies in a suitable
categorical formalism. This was first proposed by Grothendieck, who
used the term Galois--Poincar\'e categories (or rigid tensor
categories), and was then developed by Saavedra \cite{Saa}, who
introduced the now currently adopted terminology of Tannakian
categories, and by Deligne--Milne \cite{DeMi} (\cf also the more
recent \cite{De2}).

\smallskip

It is well known that there are many deep analogies between the theory
of coverings of topological spaces and Galois theory. The analogy
starts with the observation that, in cases where
the covering spaces are defined by algebraic equations, the Galois
symmetries of the equation actually correspond to deck transformations of
the covering space.

\smallskip

Grothendieck brought this initial simple analogy to far reaching
consequences. He developed a common formalism where fundamental
groups (of a space, a scheme, or much more generally a topos) and
Galois groups both fit naturally. The idea is that, in this very
general setting, the group always arises as the group of
automorphisms of a fiber functor on a suitable ``category of
coverings''. The theory of the (pro-finite) fundamental groups is
based on the existence of a fiber functor from a certain category
$\sC$ of finite \'etale covers of a connected scheme $S$, with
values in finite sets. Then such functor $\omega$ yields an
equivalence of categories between $\sC$ and $G$-sets for
$G=\Aut(\omega)$ a pro-finite group. This yields a profinite
completion of the fundamental group. For $S=\Spec(K)$, it gives
Galois theory, thus effectively bringing fundamental groups and
Galois groups within the same general formalism.

\smallskip

This is the fundamental idea that guided the development of a motivic
Galois theory. The latter appeared as a ``linear'' version of the general
formalism described above, where the fiber functor is a faithful and
exact tensor functor with values in vector spaces (instead of finite
sets), and the Galois group is the affine group scheme
$G=\Aut^\otimes(\omega)$.

\smallskip

More precisely, an abelian category is a category to which the
tools of homological algebra apply, that is, a category where the
sets of morphisms are abelian groups, there are products and
coproducts,  kernels and cokernels always exist and satisfy the
same basic rules as in the category of modules over a ring. A
tensor category over a field $k$ of characteristic zero is a
$k$-linear abelian category $\T$ endowed with a tensor functor
$\otimes: \T\times \T \to \T$ satisfying associativity and
commutativity laws defined by functorial isomorphisms, and with a
unit object. Moreover, for each object $X$, there exists a dual
$X^\vee$ and maps $ev: X\otimes X^\vee \to 1$ and $\delta: 1 \to X
\otimes X^\vee$, such that the composites $(ev\otimes 1) \circ
(1\otimes \delta)$ and $(1\otimes ev) \circ (\delta \otimes 1)$
are the identity, with an identification $k \simeq \End(1)$.

\smallskip

A {\em Tannakian category} $\T$ over $k$ is a tensor category
endowed with a fiber functor, namely a functor $\omega$ to finite
dimensional vector spaces ${\rm Vect}_K$, for $K$ an extension of
$k$, satisfying $\omega(X)\otimes \omega(Y)\simeq \omega(X\otimes
Y)$ compatibly with associativity commutativity and unit. (A more
general formulation can be given with values in locally free
sheaves over a scheme, see \cite{De2}). A {\em neutral} Tannakian
category $\T$ has a ${\rm Vect_k}$-valued fiber functor $\omega$.
In this case, the main result is that the fiber functor $\omega$
induces an equivalence of categories between $\T$ and the category
${\rm Rep}_G$ of finite dimensional linear representations of a
uniquely determined affine group scheme $G=\Aut^\otimes(\omega)$,
given by the automorphisms of the fiber functor.

\smallskip

A $k$-linear abelian category $\T$ is semi-simple if there exists
$A\subset Ob(\T)$ such that all objects $X$ in $A$ are simple (namely
$\Hom(X,X)\simeq k$), with $\Hom(X,Y)=0$ for $X\neq Y$ in $A$, and such
that every object of $\T$ is isomorphic to a direct sum of objects
in A. The affine group scheme $G$ of a neutral Tannakian category
is pro-reductive if and only if the category is semi-simple.

\smallskip

As an example, one can consider the category of finite dimensional
complex linear representations of a group. It is not hard to see
what is in this case the structure of neutral Tannakian category,
with fiber functor the forgetful functor to complex vector spaces.
The affine group scheme determined by this neutral Tannakian
category is called the ``algebraic hull" of the group. In the case
of the group $\Z$, the algebraic hull is an extension of $\hat\Z$,
with the corresponding commutative Hopf algebra given by $\Hc=\C
[e(q),t]$, for $q\in \C/\Z$, with the relations
$e(q_1+q_2)=e(q_1)e(q_2)$ and the coproduct
$\Delta(e(q))=e(q)\otimes e(q)$ and $\Delta(t)=t\otimes 1 + 1
\otimes t$.

\smallskip

The non-neutral case where $\omega$ takes values in ${\rm Vect}_K$
for some extension of $k$, or the more general case of locally
free sheaves over a scheme, can also be identified with a category
of representations, but now the group $G$ is replaced by a
groupoid (Grothendieck's Galois--Poincar\'e groupoid). This
corresponds to the fact that, even in the original case of
fundamental groups of topological spaces, it is more natural to
work with the notion of fundamental groupoid, rather than with the
base point dependent fundamental group. For our purposes, however,
it will be sufficient to work with the more restrictive notion of
neutral Tannakian category.

\medskip
\subsection*{Gauge groups and categories}\hfill\medskip

In \cite{De2}, \S 7, Deligne gives a characterization of Tannakian
categories, over a field $k$ of characteristic zero, as tensor
categories where the dimensions are positive integers. The
dimension of $X\in \T$ is defined in this context as $\Tr(1_X)$,
where $\Tr(f) =ev\circ \delta\,(f)$.

\smallskip

This characterization is very close to results developed via
different techniques by Doplicher and Roberts in the context of
algebraic quantum field theory, \cite{DR}. Their motivation was to
derive the existence of a global compact gauge group, given the
local observables of the theory. The group is obtained from a
monoidal $C^*$-category where the objects are endomorphisms of
certain unital $C^*$-algebras and the arrows are intertwining
operators between these endomorphisms. They obtain a
characterization of those monoidal $C^*$-categories that are
equivalent to the category of finite dimensional continuous
unitary representations of a compact group, unique up to
isomorphism. Though the context and the techniques employed in the
proof are different, the result has a flavor similar to the
relation between Tannakian categories and affine group schemes. In
their proof, a characterization analogous to the one of
\cite{De2}, \S 7 of the integer dimensions also plays an
important role.

\medskip
\subsection*{Pure and mixed motives}\hfill\medskip

The first constructions of a category of motives proposed by
Grothendieck covers the case of
smooth projective varieties. The corresponding motives form a
$\Q$-linear abelian category $\cM_{pure}(k)$ of {\em pure motives}.
There is a contravariant functor assigning a
motive to a variety
\begin{equation}\label{hX}
 X \mapsto h(X)=\oplus_i h^i(X).
\end{equation}
If $h^j=0$, for all $j\neq i$, the motive is {\em pure of weight $i$}.
This way a pure motive can be thought of as a ``direct summand of
an algebraic variety''. The morphisms $\Hom(X,Y)$ in the category of
motives are given by {\em correspondences}, namely algebraic cycles in
the product $X\times Y$ of codimension equal to the dimension of $X$,
modulo a suitable equivalence relation. Different
choices of the notion of equivalence for algebraic cycles produce
variants of the theory, ranging from the coarsest numerical
equivalence to the finest rational equivalence (Chow groups).
The objects of the category also include kernels of projectors,
namely of idempotents in $\Hom(X,Y)$. Thus, for $p=p^2\in \End(X)$ and
$q=q^2\in \End(Y)$, one takes $\Hom((X,p),(Y,q))=q \Hom(X,Y) p$.

\smallskip

One also adds to the objects the Tate motive
$\Q(1)$, which is the inverse of $h^2(\P^1)$. This is a pure motive of
weight $-2$. The category is endowed with a tensor product $\otimes$
and a unit $\Q(0)=h(pt)$. The Tate objects $\Q(n)$ satisfy the rule
$\Q(n+m)\cong \Q(n)\otimes \Q(m)$.

\smallskip

Grothendieck formulated a set of {\em standard conjectures} about pure
motives, which are at present still unproven. Assuming the standard
conjectures, the category of pure motives is a neutral Tannakian
category, with fiber functors given by Betti cohomology
(characteristic zero case). Thus, the category of pure motives is
equivalent to the category of representations $Rep_G$ of an affine
group scheme $G$. This group is called the {\em motivic Galois
group}. The category of pure motives is conjecturally semi-simple,
hence for pure motives $G$ is pro-reductive.

\smallskip

When one considers certain subcategories of the category of motives,
one obtains a corresponding Galois group, which is a quotient of
the original $G$. For instance, if the subcategory is generated by
a single $X$, one obtains a quotient $G_X$,
whose identity component is the Mumford--Tate group of $X$.
The subcategory of pure Tate motives, generated by
$\Q(1)$ has as motivic Galois group the multiplicative group $\bG_m$.

\smallskip

Some of the first unconditional results about motives were obtained in
\cite{Man2}. In general, a serious technical obstacle in the
development of the theory of motives, which accounts for the fact
that, decades after its conception, the theory is still largely
depending on conjectures, is
the fact that not enough is known about algebraic cycles.
The situation gets even more complicated when one wishes to consider more
general algebraic varieties, which need not be smooth projective. This
leads to the notion of {\em mixed motives} with $\M_{pure}(k)\subset
\M_{mix}(k)$.

\smallskip

Over a field of characteristic zero (where one has resolution of
singularities), one can always write such $X$ as a disjoint union
of $X_i - D_i$, where the $X_i$ are smooth projective and the
$D_i$ are lower dimensional. Thus, one can assign to $X$ a virtual
object in a suitable Grothendieck group of algebraic varieties;
however, if one wants a theory that satisfies the main
requirements of a category of motives, including the fact of
providing a universal cohomology theory (via the Ext functors),
the construction of such a category of mixed motives remains a
difficult task.

\smallskip

The main properties for a category of mixed motives
are that it should be a $\Q$-linear tensor category
containing the Tate objects $\Q(n)$ with the usual
properties, endowed with a functor $X\mapsto h(X)$ that assigns
motives to algebraic varieties, with properties like K\"unneth
isomorphisms. Moreover, the Ext functors in this category of
mixed motives define a ``motivic cohomology''
\begin{equation}\label{motH}
 E^{i,j}_2 = \Ext^i (\Q(0),h^j(X)\otimes \Q(n)) \Rightarrow  H^{i+j}_{mot}(X,\Q(n)).
\end{equation}
One expects also this motivic cohomology to come endowed with Chern
classes from algebraic $K$-theory. In fact, if one uses the decomposition
$K_n(X)\otimes \Q = \oplus_j K_n(X)^{(j)}$, where the Adams operation
$\Psi_k$ acts on $K_n(X)^{(j)}$ as $k^j$, then one expects
isomorphisms given by Chern classes
$$ ch^j: K_n(X)^{(j)} \stackrel{\simeq}{\to}
H_{mot}^{2j-n}(X,\Q(j)). $$
Such motivic cohomology will be universal with respect to all
cohomology theories for algebraic varieties
satisfying certain natural properties (Bloch--Ogus axioms).
Namely, for any such cohomology $H^*(\cdot ,\Gamma(*))$ there
will be a natural transformation
$H^*_{mot}(\cdot, \Z(*)) \to H^*(\cdot,\Gamma(*))$, compatible with
the above isomorphisms.
Mixed motives have increasing weight
filtrations preserved by the realizations to cohomology
theories.
More generally, instead of working over a
field $k$, one can consider a category $\M_{mix}(S)$ of motives (or
``motivic sheaves'') over a scheme $S$. In this case, the functors
above are natural in $S$ and to a map of schemes $f: S_1 \to S_2$ there
correspond functors $f^*$, $f_*$, $f^{!}$, $f_{!}$, behaving like the
corresponding functors of sheaves.

\smallskip

The motivic Galois group for mixed motives will then be an extension of the
pro-reductive motivic Galois group of pure motives by a pro-unipotent
group. The pro-unipotent property reflects the presence of the weight
filtration on mixed motives.

\smallskip

Though, at present, there is not yet a general construction of
such a category of mixed motives $\M_{mix}(S)$, there are constructions
of a triangulated tensor category $\cD\M(S)$, which has the right
properties to be the bounded derived category of the category
of mixed motives. The constructions of $\cD\M(S)$ due
to Levine \cite{Le} and Voevodsky \cite{Vo} are known to be
equivalent. In general, given a construction of a triangulated tensor
category, one can extract from it an abelian category by considering
the {\em heart of a $t$-structure}. A caveat with this procedure is
that it is not always the case that the given triangulated tensor
category is in fact the bounded derived category of the heart of a
$t$-structure. The available constructions, in any case, are obtained
via this general procedure of $t$-structures developed in \cite{BBD},
which can be summarized as follows.
A triangulated category ${\mathcal D}$ is an additive category
with an automorphism $T$ and a family of distinguished
triangles $X \to Y \to Z \to T(X)$, satisfying suitable axioms (which
we do not recall here). We use the notation
${\mathcal D}^{\geq n} = {\mathcal D}^{\geq 0}[-n]$ and
${\mathcal D}^{\leq n} = {\mathcal D}^{\leq 0}[-n]$, with
$X[n]=T^n(X)$ and $f[n]=T^n(f)$.
A $t$-structure consists of two full subcategories
${\mathcal D}^{\leq 0}$ and ${\mathcal D}^{\geq 0}$ with the properties:
${\mathcal D}^{\leq -1} \subset {\mathcal D}^{\leq 0}$ and
${\mathcal D}^{\geq 1} \subset {\mathcal D}^{\geq 0}$; for all
$X\in {\mathcal D}^{\leq 0}$ and all $Y\in {\mathcal D}^{\geq 1}$ one has
$\Hom_{\mathcal D} (X,Y)=0$; for all $Y\in {\mathcal D}$ there exists a
distinguished triangle as above with $X\in {\mathcal D}^{\leq 0}$
and $Z\in {\mathcal D}^{\geq 1}$.
The heart of the t-structure is the
full subcategory ${\mathcal D}^0= {\mathcal D}^{\leq 0}\cap
{\mathcal D}^{\geq 0}$. It is an abelian category. This type of
construction may be familiar to physicists in the context of mirror
symmetry, where continuous families of hearts of $t$-structures play a
role in \cite{Dou}.

\smallskip

For our purposes, we will be mostly interested in the full
subcategory of Tate motives.
The triangulated category of {\em mixed Tate motives} $\sD\M\sT(S)$
is then defined as the full triangulated subcategory of $\sD\M(S)$
generated by the Tate objects. It is possible to define on it a
$t$-structure whose heart gives a category of mixed Tate motives
$\M\sT_{mix}(S)$, provided the Beilinson--Soul\'e
vanishing conjecture holds, namely when
\begin{equation}\label{BSconj}
\Hom^j(\Q(0),\Q(n))=0, \ \ \ \text{ for } n>0, j\leq 0.
\end{equation}
where $\Hom^j(M,N)=\Hom(M,N[j])$. The conjecture \eqref{BSconj}
is known to hold in the case of a number field, where one has
\begin{equation}\label{ExtK}
\Ext^1(\Q(0),\Q(n))= K_{2n-1}(k)\otimes \Q
\end{equation}
and $\Ext^2(\Q(0),\Q(n))=0$.
Thus, in this case it is possible to extract from the triangulated tensor
category a Tannakian category $\M\sT_{mix}(k)$ of mixed Tate motives,
with fiber functor $\omega$ to $\Z$-graded $\Q$-vector spaces, $M
\mapsto \omega(M)=\oplus_n \omega_n(M)$ with
\begin{equation}\label{omegaGr}
\omega_n(M)=\Hom(\Q(n),\Gr_{-2n}^w(M)),
\end{equation}
where $\Gr_{-2n}^w(M)=W_{-2n}(M)/W_{-2(n+1)}(M)$ is the graded
structure associated to the finite increasing weight filtration
$W$.

\smallskip

The motivic Galois group of the category $\M\sT_{mix}(k)$ is then an
extension $G=U\rtimes \bG_m$, where the reductive piece is $\bG_m$ as in
the case of pure Tate motives, while $U$ is pro-unipotent. By the
results of Goncharov (see \cite{Gon}, \cite{dg}), it is known that the
pro-unipotent affine group scheme $U$ corresponds to a graded Lie
algebra ${\rm Lie}\, (U)$ that is free with one generator in each odd
degree $n\leq -3$.

\smallskip

A similar construction is possible in the case of the category
$\M\sT_{mix}(S)$, where the scheme $S$ is the set of
$V$-integers $\O_V$ of a number field $k$, for $V$ a set of finite
places of $k$. In this case, objects of $\M\sT_{mix}(\O_V)$ are
mixed Tate motives over $k$ that are unramified at each finite place
$v\notin V$. For $\M\sT_{mix}(\O_V)$ we have
\begin{equation}\label{ExtOS}
\Ext^1(\Q(0),\Q(n))=\left\{ \begin{array}{ll} K_{2n-1}(k)\otimes
\Q & n\geq 2 \\[2mm] \O_V^*\otimes \Q & n=1 \\[2mm] 0 & n\leq 0.
\end{array} \right.
\end{equation}
and $\Ext^2(\Q(0),\Q(n))=0$.
In fact, the difference between the Ext in $\M\sT_{mix}(\O_V)$ of
\eqref{ExtOS} and the Ext in $\M\sT_{mix}(k)$ of
\eqref{ExtK} is the $\Ext^1(\Q(0),\Q(1))$ which is
finite dimensional in \eqref{ExtOS} and
infinite dimensional in \eqref{ExtK}.
The category $\M\sT_{mix}(\O_V)$ is also a neutral Tannakian category,
and the fiber functor determines an equivalence of categories
between $\M\sT_{mix}(\O_V)$ and
finite dimensional linear representations of an affine group scheme
of the form $U  \rtimes \bG_m$ with $U$ pro-unipotent.
The Lie algebra $\Lie(U)$
is freely generated by a set of homogeneous generators in degree
$n$ identified with a basis of the dual of $\Ext^1(\Q(0),\Q(n))$
(\cf Prop. 2.3 of \cite{dg}). There is however no {\em canonical}
identification between $\Lie(U)$ and the free Lie algebra
generated by the graded vector space $\oplus
\Ext^1(\Q(0),\Q(n))^\vee$.

\smallskip

We mention the following case,
which will be the one most relevant in the context of perturbative
renormalization.

\begin{prop}\label{SNmotives} {\em (\cite{dg}, \cite{Gon})}
Consider the scheme $S_N=\O[1/N]$ for $k=\Q(\zeta_N)$ the
cyclotomic field of level $N$ and $\O$ its ring of integers.
For $N=3$ or $4$, the motivic Galois group of the category
$\M\sT_{mix}(S_N)$ is of the form $U  \rtimes \bG_m$, where
the Lie algebra $\Lie(U)$ is (noncanonically)
isomorphic to the free Lie algebra with one generator $e_n$ in each
degree $n\leq -1$.
\end{prop}

\section{Hilbert's XXI problem and the Riemann--Hilbert
correspondence}

Consider an algebraic linear ordinary
 differential equation, in the form of a system of rank $n$
 \begin{equation}\label{ODE}
 \frac{d}{dz} f(z) + A(z) f(z) =0
 \end{equation}
 on some open set $U=
 \bP^1(\C)\smallsetminus \{ a_1,\ldots a_r \}$,
 where $A(z)$ is an $n\times n$ matrix of rational
 functions on $U$. In particular, this includes the case of a
 linear scalar $n$th order differential equation.
The space ${\mathcal S}$ of germs of holomorphic solutions of
 \eqref{ODE} at a point $z_0\in U$ is an $n$-dimensional complex
 vector space. Moreover, given any element $\ell \in
 \pi_1(U,z_0)$, analytic continuation along a loop representing the
 homotopy class $\ell$ defines a linear automorphism of
 ${\mathcal S}$, which only depends on the homotopy class $\ell$.
 This defines the {\em monodromy representation} $\rho:
 \pi_1(U,z_0) \to {\rm Aut}({\mathcal S})$ of the differential
 system \eqref{ODE}.
A slightly different formulation requires not the {\em Fuchsian
condition} ($A(z)$ has simple poles) but the weaker {\em regular
singular condition} for \eqref{ODE}. The regularity condition at a
singular point $a_i \in \bP^1(\C)\smallsetminus U$ is a growth
condition on the solutions, namely all solutions in any strict
angular sector centered at $a_i$ have at most polynomial growth in
$1/|z-a_i|$. The system \eqref{ODE} is regular singular if every
$a_i \in \bP^1(\C)\smallsetminus U$ is a regular singular point.
The Hilbert 21st problem (or Riemann--Hilbert problem) asks
whether any finite dimensional complex linear representation of
$\pi_1(U,z_0)$ is the monodromy representation of a differential
system \eqref{ODE} with regular (or Fuchsian) singularities at the
points of $\bP^1(\C)\smallsetminus U$. A solution to the Hilbert
21st problem in the regular singular case is given by Plemelj's
theorem (\cf \cite{AnBol} \S 3). The argument
 first produces a system with the assigned monodromy on
 $U$, where in principle an analytic solution has no constraint on the
 behavior at the singularities.
 Then, one restricts to a {\em local problem} in small
 punctured disks $\Delta^*$ around each of the singularities, for
which a system
 exists with the prescribed behavior of solutions at the origin. The
 global trivialization of the holomorphic bundle on $U$ determined by
 the monodromy datum yields the
 patching of these local problems that produces a global
 solution with the correct growth condition at the singularities.

\medskip
\subsection*{From problem to correspondence}\hfill\medskip

A modern revival of interest in Fuchsian differential equation,
with a new algebraic viewpoint that slowly transformed the
original Riemann--Hilbert problem into the broad landscape of the
Riemann--Hilbert correspondence, was pioneered in the early 1960s
by the influential paper of Yuri Manin \cite{Man} on Fuchsian
modules. This new perspective influenced the work of Deligne
\cite{De1} in 1970, who solved the Riemann--Hilbert problem for
regular singular equations on an arbitrary smooth projective
variety. In this viewpoint, if $X$ is a smooth projective variety
and $U$ is a Zariski open set, with $X\smallsetminus U$ a union of
divisors with normal crossing, the data of an algebraic
differential system are given by a pair $(M,\nabla)$ of a locally
free coherent sheaf on $U$ with a connection $\nabla : M \to
M\otimes \Omega^1_{U/\C}$, while the regular singular condition
says that there exists an algebraic extension $(\bar M,
\bar\nabla)$ of the data $(M,\nabla)$ to $X$, where the extended
connection $\bar\nabla : \bar M \to \bar M \otimes
\Omega^1_{X/\C}(\log D)$ has log singularities at the divisor $D$.
The reconstruction argument for algebraic linear differential
systems with regular singularities in terms of their monodromy
representation consists then of first producing an analytic
solution $(M,\nabla)$ on $U$ with the prescribed monodromy and
then restricting to a local problem in punctured polydisks
$\Delta^*$ around the singularities, to obtain a local extension
of the form $H(z) \prod_j z_j^{B_j}$, where $H\in
\GL_n(\O_{\Delta^*})$ and the $B_j$ are commuting matrices that
give the local monodromy representation $\exp(2\pi i B_j)$ of
$\pi_1(\Delta^*)$. An important point of the argument is to show
that these local extensions can be patched together. The patching
problem does not arise when $\dim U=1$, since in that case the
divisor $D$ consists of isolated points. The construction is then
completed by showing that the global analytic extension $(\bar M,
\bar \nabla)$ obtained this way on $X$ is equivalent to an
algebraic extension.

\smallskip

Starting with the early 1980s, with  the work of Mebkhout \cite{Me1}
\cite{Me2} and of Kashiwara \cite{Ka1} \cite{Ka2}, and with the
development of the theory of perverse sheaves by Beilinson,
Bernstein, Deligne, and Gabber \cite{BBD},
the Riemann--Hilbert correspondence was
recast in terms of an equivalence of derived categories between
regular holonomic $\Dc_X$-module and perverse sheaves. A reason
for introducing the language of $\Dc$-modules (\cf
\eg \cite{GeMa} \S 8 or \cite{LM} for an overview) is that this
captures more information in a differential system
$(M,\nabla)$, than what was possible with the previous
formulations. For instance, the data
$(M,\nabla)$ fit into a de Rham type complex. Also, one may want to
consider different classes of solutions (smooth, holomorphic,
distributional, etc). This type of extra information is taken care of
by the formalism of $\Dc$-modules. Namely,
a differential equation determines a module $\Mc$ over $\Dc_X$
(differential operators on $X$ with holomorphic coefficients), with
solutions to the equation given
by $\Hom_{\Dc_X}(\Mc,\Oc_X)$. One can alter the type of solutions by
replacing $\Oc_X$ by another module $\Nc$ over $\Dc_X$, and account
for the extra structure in the data $(M,\nabla)$ by considering the de
Rham complex $\Mc \otimes_{\Oc_X} \Omega_X$. The condition of regular
singularities can be extended to modules $\Mc$ subject to another
`growth' condition, related to the module structure, compatibly with a
natural filtration of $\Dc_X$ (holonomic $\Dc$-modules). Then the
equivalence of categories extends to an equivalence of derived
categories, between regular holonomic $\Dc_X$-module and perverse
sheaves.

\smallskip

With the regular singular hypothesis replaced by the stronger
Fuchsian condition, as in Hilbert's original formulation,
counterexamples to the Riemann--Hilbert problem were later found by
Bolibruch \cite{Bol}, in the simplest case of $X=\P^1(\C)$.
On the other hand, one can instead relax the regular singular
condition and look at classes of differential systems with irregular
singularities. It is immediately clear that finite
dimensional complex linear representations of the fundamental group no
longer suffice to distinguish equations that can have very different
analytic behavior at the singularities and equal monodromy. One can
see this in a simple example, where all equations of the form
$\frac{d}{dz} f(z) + \frac{1}{z^2} P\left( \frac{1}{z} \right)f(z)
=0$ have trivial monodromy, for any polynomial $P$, but they all have
inequivalent behavior at the singularity $z=0$.

\smallskip

Thus, one needs a refinement of the fundamental group, whose
finite dimensional linear representations are equivalent to (a given
class of) irregular differential systems. There are different
approaches to the irregular case. Since we are directly interested in
the case relevant to perturbative renormalization, we might as well
restrict our attention to the one dimensional setting, namely where
$\dim U=1$ and $X$ is a compact Riemann surface. In fact, in our case
$X=\P^1(\C)$ will be sufficient, as we will be interested only in the
local problem in a punctured disk $\Delta^*$. As we discuss in
Section \ref{GalSect} below, in physical terms
$\Delta^*$ represents the space of complexified dimensions around a given
integer dimension $D$ at which the Feynman integrals of the specified
theory $\sT$ are divergent.

\smallskip

In this context, the theory that best fits our needs for the application
to renormalization was developed by Martinet and Ramis \cite{MR}, where
instead of the usual fundamental group one considers representations
of a {\em wild fundamental group}, which
arises from the asymptotic theory of divergent series and differential
Galois theory.

\medskip
\subsection*{Differential Galois theory and the wild fundamental
group}\label{DiffGalSect}\hfill\medskip

We consider a local version of the irregular Riemann--Hilbert
correspondence, in a small punctured
disk $\Delta^*$ in the complex plane around a singularity $z=0$.
We work in the context of differential Galois theory (\cf
\cite{vPS}, \cite{vP}). In this setting, one
works over a differential field $(K,\delta)$, such that the field of
constants $k=\Ker(\delta)$ is an algebraically closed field of
characteristic zero. One considers differential systems of the form
$\delta f = Af$, for some $A\in {\rm End}(n, K)$.

\smallskip

For $k=\C$, at the formal level we are then working over the
differential field of formal complex Laurent series
$K=\C((z))=\C[[z]][z^{-1}]$, with
differentiation $\delta=z\frac{d}{dz}$, while at the non-formal level
one considers the subfield $K=\C(\{z\})$ of convergent Laurent series.

\smallskip

Given a differential system $\delta f = Af$, its {\em Picard--Vessiot
ring} is a $K$-algebra with a differentiation extending $\delta$. As a
differential algebra it is simple and is generated over $K$ by the
entries and the inverse determinant of a fundamental matrix for the
equation $\delta f = Af$. The {\em differential Galois group} of the
differential system is given by the automorphisms of the
Picard--Vessiot ring commuting with $\delta$.

\smallskip

The formalism of Tannakian categories, that we discussed in Section
\ref{MotSect} in the context of motives, reappears in the present
context and allows for a description of the differential Galois group
that fits in the same general picture we recalled regarding motivic
Galois groups.

\smallskip

In fact, if we consider the set of all possible such differential systems
(differential modules over $K$), these form a neutral
Tannakian category, which can therefore be identified with the
category of finite dimensional linear
representations of a unique affine group scheme over the field $k$.

\smallskip

Similarly to what we discussed in the case of motivic Galois groups,
any subcategory $\T$ that inherits the structure of a neutral
Tannakian category in turn corresponds to an affine group scheme $G_\T$.
This is the universal differential Galois group of the class of
differential systems that form the category $\T$. It can be realized
as the automorphisms group of the universal Picard--Vessiot ring
$R_\T$. The latter is generated over $K$ by the entries and inverse
determinants of the fundamental matrices of all the differential
systems considered in the category $\T$.

\smallskip

There is therefore a clear analogy between the induced motivic Galois
groups of certain subcategories of, say, the category of mixed Tate
motives that we discussed in Section \ref{MotSect}, and the
differential Galois group of certain classes of differential
systems defining subcategories of the neutral Tannakian category of
irregular differential systems over a differential field $K$. Our main
result of \cite{cmln}, which we discuss in Section \ref{GalSect}
below, shows that the theory of perturbative renormalization (in the
DimReg and minimal subtraction scheme)
identifies a class of differential systems (dictated by physical
assumptions), whose differential Galois group is the motivic Galois
group of Proposition \ref{SNmotives}.

\smallskip

The regular--singular case can be seen in this context as follows.
The subcategory of regular--singular differential modules over
$K=\C((z))$ is a neutral Tannakian category equivalent to $Rep_G$,
where the affine group scheme $G$
is the algebraic hull $\bar\Z$ of $\Z$, generated by the formal monodromy
$\gamma$. The latter is the automorphism of the universal
Picard--Vessiot ring acting by
$\gamma \, Z^a = \exp(2\pi i a)\, Z^a$ and $\gamma\, L = L +
2\pi i$ on the generators
$\{ Z^a \}_{a\in \C}$ and $L$, which correspond, in angular sectors,
to the powers $z^a$ and the function
$\log(z)$ (\cf \cite{vPS} \S III, \cite{vP}).

\smallskip

When one allows for an arbitrary degree of irregularity for the
differential systems $\delta f =Af$, the universal Picard--Vessiot
ring of the formal theory $K=\C((z))$ is generated by elements $\{ Z^a
\}_{a\in \C}$ and $L$ as before, and by additional elements $\{ E(q)
\}_{q\in \Ec}$,
where $\Ec= \cup_{\nu\in \N^\times} \Ec_\nu$, for $\Ec_\nu=
z^{-1/\nu} \C [ z^{-1/\nu} ]$.
These additional generators correspond, in local sectors, to functions of
the form $\exp(\int q\, \frac{dz}{z})$ and satisfy
relations $E(q_1+q_2)=E(q_1)E(q_2)$ and $\delta E(q)=q E(q)$.

\smallskip

Correspondingly, the universal
differential Galois group $\cG$ is described by an extension
${\mathcal T}\rtimes \bar\Z$, where ${\mathcal T}=\Hom(\Ec, \C^*)$ is
the {\em Ramis exponential torus}. The algebraic hull
$\bar\Z$ generated by the formal monodromy $\gamma$ acts
as an automorphism of the universal Picard--Vessiot ring by
the same action as above on the $Z^a$ and on
$L$, and by $\gamma\, E(q) = E(\gamma q)$ on the
additional generators, where the action on $\Ec$ is given by the
 action $ \gamma : q( z^{-1/\nu}) \mapsto q(
 \exp( -2\pi i/\nu) \, z^{-1/\nu})$
of $\Z/\nu\Z$ on $\Ec_\nu$.
The exponential torus acts by automorphisms of the
universal Picard--Vessiot ring $\tau\,Z^a=Z^a$, $\tau\, L=L$ and
$\tau\, E(q)=\tau(q) E(q)$, and the formal monodromy acts on the
exponential torus by $(\gamma\tau)(q)=\tau(\gamma q)$.

\smallskip

Thus, at the formal level, the local irregular Riemann--Hilbert
correspondence establishes an equivalence of categories between
the differential modules over $K=\C((z))$ and finite dimensional
linear representations of $G={\mathcal T}\rtimes \bar\Z$. Ramis'
wild fundamental group \cite{MR} further extends this irregular
Riemann--Hilbert correspondence to the non-formal setting. In
general, when passing to the non-formal level over convergent
Laurent series $K=\C(\{z\})$, the universal differential Galois
group acquires additional generators, which depend upon
resummation of divergent series and are related to the Stokes
phenomenon. However, there are specific classes of differential
systems (subcategories of differential modules over $K$), for
which the differential Galois group is the same over $\C((z))$ and
over $\C(\{z\})$ (\cf \eg Proposition 3.40 of \cite{vPS}). In such
cases, the wild fundamental group consists only of the exponential
torus and the formal monodromy. This is, in fact, the case in the
class of differential systems we obtain from the theory of
perturbative renormalization, hence we do not need to discuss here
the more complicated case where Stokes phenomena are present, and
we simply refer the interested reader to \cite{MR}, \cite{vPS},
and \cite{vP}.

\section{Cartier's dream of a cosmic Galois group}\label{GalSect}

In the section ``I have a dream'' of \cite{Cart}, Pierre Cartier
formulated the hypothesis of the existence of a ``cosmic Galois
group'', closely related to the Grothendieck--Teichm\"uller group
\cite{Gro2}, underlying the Connes--Kreimer theory of perturbative
renormalization, that would relate quantum field theory to the theory
of motives and multiple zeta values.

\smallskip

We present in this section the main result of \cite{cmln}, which
realizes Cartier's suggestion, by reformulating the Connes--Kreimer
theory of perturbative renormalization in the form of a suitable
Riemann--Hilbert correspondence.

\subsection*{Equisingular connections}\hfill\medskip

The first step, in order to pass to this type of geometric formulation, is to
identify the loops $\gamma_\mu(z)=\gamma_{\mu,-}(z)^{-1}
\gamma_{\mu,+}(z)$ with solutions of suitable differential
equations. The idea of reformulating a Birkhoff factorization
problem in terms of a class of differential equations is familiar to
the analytic approach to the Riemann--Hilbert problem (\cf
\cite{Boj}). In our setting, the key that allows us to pass from
the Birkhoff factorization
to an appropriate class of differential systems is provided by the `t
Hooft relations in the form of Proposition \ref{thooftrel}
and the scattering formula \eqref{scattering}, reformulated more
explicitly in terms of iterated integrals.

\smallskip

Here the main tool is the {\em time ordered exponential}, formulated
mathematically in terms of Chen's iterated integrals \cite{Chen1},
\cite{Chen2}, also known (in the operator algebra context) as Araki's
expansional \cite{Araki}.

\smallskip

We consider a commutative Hopf algebra $\cH$
that is graded in positive degree and connected, with $G$ the
corresponding affine group scheme and $\fg=\Lie G$.
We assume that $\Hc$ is, in each degree, a finite dimensional vector
space. Given a $\fg(\C)$-valued smooth
function $\a(t)$ where $t\in[a,b]\subset \R$ is a real parameter, the
expansional is defined by the expression
\begin{equation}\label{expansional}
 {\bf {\rm T}e^{\int_a^b\,\a(t)\,dt}}=\,1+\, \sum_1^\infty \int_{a\leq
s_1\leq \cdots\leq
s_n\leq b} \,\a(s_1)\cdots\,\a(s_n) \prod ds_j \,,
\end{equation}
where the products are in the dual algebra $\Hc^*$ and $1\in \Hc^*$ is
the unit given by the augmentation $\ve$. When paired with any element
$x\in \Hc$, \eqref{expansional} reduces to a finite sum, which defines
an element in $G(\C)$.

\smallskip

The fact that, when pairing with elements in
$\Hc$ one reduces to an algebraic (polynomial) case plays an important
role. In particular, it is related to the fact that, for the class of
differential systems we consider, the differential
Galois group remains the same in the formal and in the non-formal
case, and we need not take into account the possible presence of
Stokes' phenomena.

\smallskip

We are particularly interested in the following property of the
expansional: \eqref{expansional} is the value $g(b)$ at $b$ of
the unique solution $g(t)\in G(\C)$ with value $g(a)=1$ of
the differential equation
\begin{equation}\label{diffexp}
 dg(t)=\,g(t)\,\a(t)\,dt\,.
\end{equation}

\smallskip

More generally, if $(K,\delta)$ is a differential field with $K\supset
\C$ and if $g\in
G(K)$ and $g'=\delta(g)$ is the linear map $\Hc\to K$ defined as
$g'(x)=\, \delta(g(x))$ for $x\in \Hc$, then the
logarithmic derivative $D(g)$ is defined as the linear map $\Hc\to K$
of the form $D(g)=\,g^{-1}\star\, g'$, with the product dual to the
coproduct of $\cH$. It satisfies
$$
\langle D(g),x\,y\rangle=\,\langle D(g),x\rangle\,\ve(y)
+\,\ve(x)\,\langle D(g),y\rangle \qq x,y \in \Hc\,,
$$
hence it gives an element in the Lie algebra $D(g) \in \fg(K)$. We
will work here with the field of convergent Laurent series $K=\C(\{z
\})$.

\smallskip

If we consider over $\Delta^*$ a differential system of the form
\begin{equation}\label{Dfomega}
Df =\omega,
\end{equation}
where $\omega$ is a flat $\fg(K)$-valued connection, then the condition of
trivial monodromy
\begin{equation}\label{mono}
{\bf {\rm T}e^{\int_0^1\,\g^*\omega}} =1,
\end{equation}
for $\gamma\in \pi_1(\Delta^*,z_0)$, ensures the existence of a
solution. In the expansional form this is given by
\begin{equation}\label{solution}
g(z) =\;{\bf {\rm T}e^{\int_{z_0}^z\,\omega}}\,,
 \end{equation}
independently of the path in $\Delta^*$ from $z_0$ to $z$.

\smallskip

The notion of equivalence relation that we consider for differential
systems of the form \eqref{Dfomega} is the following:
two connections $\omega$ and $\omega'$ are equivalent
iff they are related by a gauge transformation $h \in  G(\O)$, with $\O
\subset  {K}$ the subalgebra of regular functions,
\begin{equation}\label{gaugetransf}
 \omega' = Dh + h^{-1} \omega\, h.
\end{equation}
The behavior of solutions at the singularity is the same for all
equivalent connections. When we regard the solutions as $G(\C)$-valued
loops, the equivalence \eqref{gaugetransf} of the connections
translates to the fact that the loops have the same negative part of
the Birkhoff decomposition.

\smallskip

So far we have not taken into account the fact that, in the case of
perturbative renormalization, the loop $\gamma_\mu(z)$ that corresponds to the
unrenormalized theory depends on
the mass parameter $\mu$, as discussed above in Section
\ref{MassSect}. Because of the presence of this parameter, the
geometric reformulation in terms of a class of differential systems
takes place, in fact, not just on the 1-dimensional (infinitesimal)
punctured disk $\Delta^*$ representing the complexified dimensions of
DimReg, but on a principal $\bG_m(\C)=\C^*$-bundle over $\Delta^*$.

\smallskip

As we discuss below, the fact that the loop $\gamma_\mu(z)$ satisfies
the properties \eqref{gammamu-} and
\eqref{thetagamma} will make it possible to treat this case, which
lives naturally over a 2-dimensional space, by applying the same
techniques described in Section
\ref{DiffGalSect} for the irregular Riemann--Hilbert correspondence
over the 1-dimensional domain $\Delta^*$.

\smallskip

The conditions \eqref{gammamu-} and
\eqref{thetagamma} determine a class of differential systems
associated to perturbative renormalization. This is given by
equivalence classes of flat {\em equisingular} $G(\C)$-connections, where
$G={\rm Difg}(\sT)$.

\smallskip

Let $\pi\,:B\to \Delta$ be a principal $\bG_m(\C)=\C^*$-bundle,
identified with $\Delta \times \C^*$ by the non-canonical choice of a
section $\sigma\,: \Delta \to B$, $\sigma(0)= y_0$. Physically, the
latter corresponds to a choice of the Planck constant. Let
$P=B\times G(\C)$ be the trival principal $G(\C)$-bundle, and $B^*$ and $P^*$
the restrictions to the punctured disk $\Delta^*$.

\smallskip

We say that the  connection $\omega$ on $P^*$ is {\em equisingular} if
it is $\bG_m$-invariant and its restrictions to sections of the
principal bundle $B$ that agree at $0\in \Delta$ are mutually
equivalent, in the sense that they are related by a gauge
transformation by a $G(\C)$-valued $\bG_m$-invariant map $h$ regular in $B$.

\smallskip

The notion of {\em equisingularity} is introduced
as a geometric reformulation of the properties \eqref{gammamu-} and
\eqref{thetagamma}. In fact, the property that, when approaching the
singular fiber, the type of singularity
does not depend on the section along
which one restricts the connection but only on the value of the
section at $0\in \Delta$ corresponds to the fact that the counterterms
are independent of the mass scale, as in \eqref{gammamu-}.

\smallskip

Thus, we have identified a class of differential systems associated to
a physical theory $\sT$, namely the equivalence classes of flat
equisingular $G(\C)$-valued connections on $P$, where $G={\rm
Difg}(\sT)$. We can then proceed to investigate the Riemann--Hilbert
correspondence underlying this class of differential systems.

\smallskip

The first step consists of writing solutions of
\eqref{gammamu-} and \eqref{thetagamma} in expansional form through
the following result, which we can view as a stronger version of
the `t Hooft relations.

\begin{prop}\label{expans}
Let $\g_\mu(z)$ be a family of $G(\C)$-valued loops
satisfying the properties \eqref{gammamu-} and
\eqref{thetagamma}.
 Then there exists a unique $\beta \in {\rm Lie}\,G(\C)$
 and a loop $\g_{\rm reg}(z)$ regular at $z=0$,
 such that
\begin{equation}\label{solexp}
  \g_\mu(z) =\,{\bf {\rm T}e^{-\frac{1}{z}\,
 \int^{-z \log\mu}_\infty\,\t_{-t}(\beta)\,dt}}\;
  \t_{z\log\mu}(\g_{\rm reg}(z))\,.
\end{equation}
Conversely, for any $\beta$ and regular loop $\g_{\rm reg}(z)$ the expression
\eqref{solexp} gives a solution to \eqref{gammamu-} and
\eqref{thetagamma}.
The Birkhoff
 decomposition of $\g_\mu(z)$ is of the form
\begin{equation}\label{gammaplusminus}
\begin{array}{ll} \g_{\mu^+}(z)= & {\bf {\rm T}e^{-\frac{1}{z}
  \,\int_0^{-z \log\mu}\,\t_{-t}(\beta)\,dt}}\;
 \t_{z\log\mu}(\g_{\rm reg}(z))\,, \\[3mm]
 \g_-(z) = & \,{\bf {\rm
 T}e^{-\frac{1}{z}\,\int_0^\infty\,\t_{-t}(\beta)\,dt}}\,.
 \end{array} \end{equation}
\end{prop}

\smallskip

Using the equivalent geometric formulation in terms of flat
equisingular connections, one then obtains the following result.

\begin{prop}\label{connbeta}
Let $\omega$ be a flat equisingular $G(\C)$-connection. There
exists a unique element $\beta \in \Lie\, G(\C)$, such that
$\omega$ is equivalent to the flat
 equisingular connection $D\g$ for
\begin{equation}\label{solexpm}
 \g(z,v) =\,{\bf {\rm T}e^{-\frac{1}{z}\,
 \int^{v}_0\,u^Y(\beta)\,\frac{du}{u}}}\;
\in G(\C)\; \,,
\end{equation}
with the integral performed on the straight path $u=t v$,
$t\in[0,1]$.
\end{prop}

\smallskip

Here a crucial point is the fact that the monodromies with respect to
the two generators of $\pi_1(B^*)$ vanish for flat equisingular
connections. As we will see in the next section, this fact will be
reflected in the form of the affine group
scheme associated to the category of equivalence classes of flat
equisingular connections (the differential Galois group), which will
only contain the part corresponding to the Ramis exponential torus and
no contribution from the monodromy. The correspondence of Proposition
\ref{connbeta} is independent of the choice of the trivialising
section $\sigma$ of $B$.

\medskip
\subsection*{The Riemann--Hilbert correspondence}\hfill\medskip

So far we have been working with an assigned quantum field theory
$\sT$ and the corresponding affine group scheme $G={\rm
Difg}(\sT)$. We now pass to considering a universal setting, which
encompasses all theories. This is achieved by considering, instead
of flat equisingular $G(\C)$-connections, the category of
equivalence classes of all {\em flat equisingular bundles}. For a
specific physical theory, the corresponding category of
equivalence classes of flat equisingular $G(\C)$-connections can
be recovered from this more general setting by considering the
subcategory of those flat equisingular bundles that are finite
dimensional linear representations of $G^*=G\rtimes \bG_m$. This
is analogous to what happens when one specializes motivic Galois
groups to sucategories of motives, or differential Galois groups
to subcategories of differential systems. We describe now in
detail the universal setting, with the corresponding group of
symmetries and the way it specializes to a given physical theory.

\smallskip

The category of equivalence classes of flat equisingular bundles
has as objects $\Theta=(E,\nabla)$ pairs of a finite dimensional
$\Z$-graded vector space $E$ and an equisingular flat
$W$-connection $\nabla$. To define the latter, we consider the
vector bundle $\tilde{E}=B\times E$ with the action of $\bG_m$
given by the grading and with the weight filtration defined by
$W^{-n}(E)=\oplus_{m\geq n} E_m$. A $W$-connection is a connection
on the restriction of $\tilde E$ to $B^*$, which is compatible
with the weight filtration and induces the trivial connection on
the associated graded. The connection $\nabla$ in the data
above is a flat $W$-connection that satisfies the equisingular
condition, that is, it is $\bG_m$-invariant and the restrictions
to sections $\sigma$ of $B$ with $\sigma(0)=y_0$ are all
$W$-equivalent on $B$, where the equivalence relation is realized
by an isomorphism of the vector bundles over $B$, compatible with
the filtration and identity on the associated graded, that
conjugates the connections. We consider the data
$\Theta=(E,\nabla)$ as $W$-equivalence classes.

\smallskip

As usual, it is a bit more delicate to define morphisms than
objects.

For a linear map $T: E\to E'$, consider the $W$-connections
$\nabla_j$, $j=1,2$, on $\tilde{E'}\oplus \tilde{E}$ of the form
\begin{equation}\label{nablaj}
\nabla_1 = \left( \begin{matrix}\nabla' &0 \cr 0 &\nabla \cr
\end{matrix} \right) \ \ \ \text{ and } \ \ \
 \nabla_2 = \left( \begin{matrix}\nabla' &T\,\nabla-\,\nabla'\,T
\cr 0 &\nabla \cr \end{matrix} \right),
\end{equation}
where $\nabla_2$ is the conjugate of $\nabla_1$ by the unipotent
matrix
$$
\left( \begin{matrix}1 &T \cr 0 &1 \cr \end{matrix}\right)\,.
$$
Morphisms $T\in{\rm Hom}(\Theta, \Theta')$ in the category of
equisingular flat bundles are linear maps $T: E\to E'$ compatible
with the grading and such that the connections $\nabla_j$ of
\eqref{nablaj} are $W$-equivalent on $B$. The condition is
independent of the choice of representatives for the connections
$\nabla$ and $\nabla'$.

\smallskip

The category $\Ec$ of equisingular flat bundles is a tensor
category over $k=\C$, with a fiber functor $\omega: \Ec \to
Vect_\C$ given by
\begin{equation}\label{omegaE}
 \omega: \Theta=(E,\nabla) \mapsto E.
\end{equation}
In fact, one can refine the construction and work over the field
$k=\Q$, since the universal singular frame (see \eqref{framecoeff} below),
in which one expresses the connections, has rational
coefficients. In this case, the fiber functor $\omega: \Ec_\Q \to
Vect_\Q$ is of the form $\omega = \oplus \,\omega_n$, with
$$
\omega_n(\Theta)=\,{\rm Hom}(\Q(n), \Gr_{-n}^W(\Theta))\,,
$$
where $\Q(n)$ denotes the object in $\Ec_\Q$ given by the class of
the pair of the trivial bundle over $B$ with fiber a
one-dimensional $\Q$-vector space placed in degree $n$ and the
trivial connection.

\smallskip

Let $\cF(1,2,3,\cdots)_{\bullet}$ be the free graded Lie algebra
generated by one element $e_{-n}$ in each degree $n\in \Z_{>0}$,
and let
\begin{equation}\label{hopfu}
\Hc_u=\; \cU(\cF(1,2,3,\cdots)_{\bullet})^\vee
\end{equation}
be the commutative Hopf algebra obtained by considering the graded
dual of the enveloping algebra $\cU(\cF)$. We can then identify
explicitly the affine group scheme associated to the neutral
Tannakian category of flat equisingular bundles as follows (\cf
\cite{cmln} \cite{CM2}).

\begin{thm}\label{EU}
The category $\Ec$ of flat equisingular bundles
is a neutral Tannakian category, with fiber
functor \eqref{omegaE}. It is equivalent to the category ${\rm
Rep}_{U^*}$ of finite dimensional linear representations of the
affine group scheme $U^*=U\rtimes \bG_m$, where $U$ is the
pro-unipotent affine group scheme associated to the Hopf algebra
$\Hc_u$ of \eqref{hopfu}.
\end{thm}

The affine group scheme $U^*$ is a motivic Galois group. In fact,
by results of Goncharov and Deligne (\cite{dg}, \cite{Gon}, see
Proposition \ref{SNmotives} above), we have the following
identification of the ``cosmic Galois group'' $U^*$.

\begin{prop} \label{motgal} There is a
(non-canonical)  isomorphism
\begin{equation}\label{MotU}
U^* \cong G_{\cM_T}(\fO) \,.
\end{equation}
of the affine group scheme $U^*$ with the motivic Galois group
$G_{\cM_T}(\fO)$ of  the scheme $S_N$ of $N$-cyclotomic integers,
for $N=3$ or $N=4$.
\end{prop}

The fact that we only have a noncanonical identification suggests
that there should be an explicit identification dictated by the
form of the iterated integrals that give the expansionals defining
the equisingular connections as in Proposition \ref{expans}. This
should be related to Kontsevich's formula for multiple zeta values
as iterated integrals generalized by Goncharov to multiple
polylogarithms ${\rm Li}_{\,k_1,\ldots,k_m}(z_1,z_2,\ldots,z_m)$,
in terms of the expansional ${\bf {\rm
T}e^{\int_0^1\,\a(z)\,dz}}$, with the connection
$$ \alpha(z)dz = \sum_{a\in  \mu_m\cup \{0\}} \, \frac{dz}{z-a} \,\, e_a. $$

Notice, moreover, that the group $U^*$, as the differential Galois
group in the formal theory of equisingular connections,
corresponds to the Ramis exponential torus. In fact, we have no
contribution from the monodromy, a fact on which the proof of
Proposition \ref{connbeta} depends essentially, and we also do not
have Stokes phenomena, hence, as far as the differential Galois
group is concerned, we can equally work in the formal or in the
non-formal setting.

\medskip
\subsection*{The renormalization group as a Galois group}\hfill\medskip

The formulation of Theorem \ref{EU} is universal with respect to
physical theories. When we consider a particular choice of a
renormalizable theory $\sT$, we restrict the category of
equisingular flat bundles to the full subcategory of finite
dimensional linear representations of $G^*=G\rtimes \bG_m$, for
$G={\rm Difg}(\sT)$. In this case, the Riemann--Hilbert
correspondence specializes to a morphism of differential Galois
groups, as follows.

\begin{prop} \label{rhoG}
Let $G$ be a positively graded pro-unipotent affine group scheme.
Then there exists a canonical bijection between equivalence
classes of flat equisingular $G(\C)$-connections and graded
representations $\rho \, : U\,\to G$, of the affine group scheme
$U$ in $G$. Compatibility with the grading implies that $\rho$
extends to a homomorphism $\rho^* \, : U^*\,\to G^*$, which is the
identity on $\bG_m$.
\end{prop}

\smallskip

This is a reformulation of the result of Proposition
\ref{connbeta}. In fact, more explicitly, the representation
$\rho$ of Proposition \ref{rhoG} is obtained as follows. We can
write an element $\beta$ in $\Lie\,G$ as an infinite formal sum
\begin{equation}\label{betasum}
\beta=\; \sum_1^\infty\;\beta_n\,,
\end{equation}
where, for each $n$, $\beta_n$ is homogeneous of degree $n$ for
the grading, \ie $Y(\beta_n)=n \beta_n$. Thus, assigning $\beta$
with the action of the grading is the same as giving a collection
of homogeneous elements $\beta_n$, with no restriction besides
$Y(\beta_n)=n \beta_n$. In particular, there is no condition on
their Lie brackets, hence assigning such data is equivalent to
giving a homomorphism from the affine group scheme $U$ to $G$, by
assigning, at the Lie algebra level, the generator $e_{-n}$ to the
component $\beta_n$ of $\beta$.

\smallskip

In particular, the result above means that we can realize the
renormalization group as a Galois group. In fact, recall that, for
an assigned theory $\sT$, the corresponding $\beta$ that
determines the counterterms $\gamma_{-}(z)$ is the infinitesimal
generator of the renormalization group \eqref{rengroup}. The
representation $\rho: U^* \to G^*$ then determines a lifting of
the renormalization group ${\bf rg}$ to a canonical 1-parameter
subgroup of $U^*$, obtained by considering the element
\begin{equation}\label{esum}
e=\; \sum_1^\infty\;e_{-n}\,,
\end{equation}
in the Lie algebra $\Lie\,U$. As $U$ is a pro-unipotent affine
group scheme, $e$ defines a morphism of affine group schemes
\begin{equation}\label{rgU}
{\bf{rg}}\; :\,\bG_a \,\to \,U\,,
\end{equation}
from the additive group $\bG_a$ to $U$.

\smallskip

Thus, the rest of the affine group scheme $U$ can be throught of
as further symmetries that refine the action of the
renormalization group on a given physical theory. More precisely,
restricting the attention to a generator $e_{-n}$ of the Lie
algebra of $U$ corresponds to considering the flow generated by
the degree $n$ component of the $\beta$ function with respect to
the grading by loop number. Thus, from a physical point of view
the Galois group $U$ accounts for a decomposition of the action of
the renormalization group in terms of a family of flows restricted
to the $n$-loops theory.

\medskip
\subsection*{Universal singular frame}\hfill\medskip

The element $e\in \Lie\,U$ defined in \eqref{esum} determines a
``universal singular frame'' given by
\begin{equation}\label{univ}
\g_U(z,v) =\,{\bf {\rm T}e^{-\frac{1}{z}\,
 \int^{v}_0\,u^Y(e)\,\frac{du}{u}}}\;
\in U\;
  \,.
\end{equation}
This is obtained by applying Proposition \ref{connbeta} to the
affine group scheme $U$. This can be expressed explicitly in terms
of iterated integrals in the form
\begin{equation}\label{framecoeff}
\g_U(z,v) =\,\sum_{n \geq
0}\,\sum_{k_j>0}\,\frac{e_{-k_1}e_{-k_2}\cdots e_{-k_n}}
{k_1\,(k_1+k_2)\cdots (k_1+k_2+\cdots +k_n)}\,v^{\sum
k_j}\,z^{-n},
\end{equation}
with $e_{-n}$ the generators of $\Lie \,U$. This expansion has
rational coefficients. The coefficients are the same as those
occurring in in the local index formula of Connes--Moscovici
\cite{cmindex}, where the renormalization group idea is used in
the case of higher poles in the dimension spectrum.

\smallskip

The Birkhoff factorization in $U$, applied to the universal singular
frame, yields universal counterterms that maps under the
representation $\rho: U \to {\rm Difg}(\sT)$ to the counterterms of a
specific theory $\sT$.

\section{Renormalization and geometry}

Quantum mechanics allows for two equivalent formulations of
physics at the macroscopic scale, based on coordinate and
momentum space, dual to one another by Fourier transform, while
gravity, relativistically formulated in terms of the geometry of
space-time, appears to privilege coordinates over momenta.

\smallskip

In the quantum theory of fields, at the perturbative level,
Feynman integrals are computed in momentum space, using the
dimensional regularization scheme. A nice historical and
motivational perspective on how this came to be the general
``accepted paradigm'' in the context of renormalizable
perturbation theory can be found in Veltman's paper \cite{Vel}. As
Veltman suggests, one can assume perturbative field theory as the
starting point, defined in terms of Feynman diagrams using
dimensional regularization (he refers to this as the ``dimensional
formulation''). This is very much the approach followed by the
Connes--Kreimer theory and by our present work, where such
physical data, taken as the given starting point, are reformulated
in a more satisfactory conceptual perspective.

\smallskip

It is also possible to follow a different approach and to
consider the problem of perturbative renormalization in
coordinate space, working geometrically in terms of
Fulton--MacPherson compactifications. A mathematical theory of
perturbative renormalization under this point of view was
developed recently by Kontsevich \cite{Kon}. It has the advantage
of introducing directly geometric objects like algebraic
varieties, hence a natural setting for an explicit action of
motivic Galois symmetries (\cf also \cite{Kon1}).

\smallskip

As stressed by Veltman \cite{Vel}, space and time do not occur at
all in the dimensional formulation, as coordinate space exists
solely as Fourier transform of momentum space, which ceases to be
defined when momentum space is continued to complex dimension.
Notions associated to coordinate space, such as length and time
measurements, must be recovered through the gravitational field,
with graviton-fermion interactions determined by gauge invariance
and Ward identities. Thus, a viewpoint that favors momentum rather
than coordinate space is necessarily closer to noncommutative
geometry than to classical algebraic geometry. In noncommutative
geometry the metric properties of space are assigned not by a
local coordinates description of the metric tensor, but through a
``dual viewpoint'', spectrally, in terms of the Dirac operator,
hence they continue to make sense on spaces that no longer exist
classically. This appears to be a promising approach to reconcile
space (no longer defined classically) with the dimensional
formulation.

\smallskip

It is important to stress, in this respect, that the formulation
of Riemannian spin geometry in the setting of noncommutative
geometry, in fact, has already built in the possibility of
considering a geometric space at dimensions that are complex
numbers rather than integers. This is seen through the dimension
spectrum, which is the set of points in the complex plane at which
a space manifests itself with a nontrivial geometry. There are
examples where the dimension spectrum contains points off the real
lines (\eg the case of Cantor sets), but here one is rather
looking for something like a deformation of the geometry in a
small neighborhood of a point of the dimension spectrum, which
would reflect dimensional regularization. The possibility of
recasting the dimensional formulation in the setting of
noncommutative geometry may prove very useful in the problem of
extending at a fully quantum level the geometric interpretation of
the standard model of elementary particle physics provided by
noncommutative geometry (\cite{CoSM}, \cite{ChCo}).

\smallskip

An important related question, which may be a starting point for
such broader program, is to understand the precise relation
between the universal singular frame and the local index formula,
which in turn may cast some new light on the issue of the relation
of the theory of perturbative renormalization illustrated here and
noncommutative geometry. Since the local index formula of
Connes--Moscovici is closely related to chiral anomalies, a direct
comparison with the local index formula will involve a well known
problem associated to dimensional regularization in the chiral
case, namely the technical issue of how to extend the definition
of the product
\begin{equation}\label{gamma5}
 \gamma_5= i \gamma^0\gamma^1\gamma^2\gamma^3,
\end{equation}
of the $\gamma$ matrices, which integer dimension $D=4$ satisfies
the Clifford relations $\{ \gamma^\mu,\gamma^\nu \} = 2
g^{\mu\nu}\, I$, with $\Tr(I)=4$, and anticommutativity $\{
\gamma_5, \gamma^\mu \} =0$. The $\gamma^5$ problem, however, is
not considered a serious obstacle to the application of
dimensional regularization, as there are good methods to address
it (\cf \cite{MSR} for a recent discussion of this issue). For
instance, the $\gamma^5$ problem is addressed by the
Breitenlohner--Maison approach, in which one does not give an
explicit expression for the gamma matrices in complex dimension,
but just defines them (and the $\gamma_5$ given by \eqref{gamma5})
through their formal properties. In \cite{DK1}, Kreimer described
another approach to the problem, in which $\gamma_5$ still
anticommutes with $\gamma^\mu$ but the trace is no longer cyclic, an
approach that is expected to be equivalent to the one of
Breitenlohner--Maison (\cf \cite{DK1}, \S 5).

\smallskip

Finally, we would like to end on a more speculative tone, by
mentioning a very different source for the idea of the existence
of a deformation of geometry to non-integral complex dimensions.
In arithmetic geometry, the Beilinson conjectures relate the
values of the first non-vanishing derivative at integer points of
the motivic $L$-functions of algebraic varieties to periods,
namely numbers obtained by integration of algebraic differential
forms on algebraic varieties (\cf \eg \cite{KoZa}). This process
of considering the expansion in a neighborhood of an integer point
is reminiscent of the procedure of Dim-Reg, where one considers
the Feynman integrals in an infinitesimal neighborhood of the
integer dimension $D$. Based on this analogy, it becomes extremely
suggestive to imagine that the Beilinson conjectures may be
related to a ``dimensional regularization of algebraic varieties
and periods'', and that there may be a geometric interpretation
even for the values at non-integer points, in terms of some
(noncommutative) geometry in complex dimension.

\end{document}